\documentclass[journal=jacsat,manuscript=article, 50pt]{achemso}

\usepackage{setspace}
\usepackage{xcolor}
\usepackage{soul}
\usepackage[font={small}]{caption}
\usepackage{graphicx} 
\usepackage[cal=pxtx]{mathalfa}
\usepackage{array}    
\usepackage{booktabs} 
\usepackage{amsmath,amssymb}  
\usepackage{geometry} 
\usepackage{comment}
\usepackage[version=3]{mhchem} 
\usepackage{amsmath}

\author{Sainath A. Barbhai}
\altaffiliation{These authors contributed equally to this work.}
\affiliation[University of Illinois at Urbana-Champaign]
{Department of Aerospace Engineering, University of Illinois at Urbana-Champaign, Urbana, Illinois 61801, USA}

\author{Zhengyu Yang}
\altaffiliation{These authors contributed equally to this work.}
\author{Jie Feng}
\email{jiefeng@illinois.edu}

\affiliation[University of Illinois at Urbana-Champaign]
{Department of Mechanical Science and Engineering, University of Illinois at Urbana-Champaign, Urbana, Illinois 61801, USA}

 \title[]
   {Effect of a polymeric compound layer on jetting dynamics produced by bursting bubbles}

\abbreviations{IR,NMR,UV}

\begin{document}

\singlespacing

\fontsize{11pt}{13.2pt}\selectfont
\begin{abstract}

Jetting dynamics from bursting bubbles play a key role in mediating mass and momentum transport across the air-liquid interface. In marine environments, this phenomenon has drawn considerable attention due to its role in releasing biochemical contaminants, such as extracellular polymeric substances, into the atmosphere through aerosol production. These biocontaminants often exhibit non-Newtonian characteristics, yet the physics of bubble bursting with a rheologically complex layer at bubble-liquid interfaces remains largely unexplored. In this study, we experimentally investigate the jetting dynamics of bubble bursting events in the presence of such polymeric compound layers. Using bubbles coated by a polyethylene oxide solution, we document the cavity collapse and jetting dynamics produced by bubble bursting. At a fixed polymer concentration, the jet velocity increases while the jet radius decreases with an increasing compound layer fraction, as a result of stronger capillary wave damping due to capillary wave separation at the compound interface as well as the formation of smaller cavity cone angles during bubble cavity collapse. These dynamics produce smaller and more numerous jet drops. Meanwhile, as the polymer concentration increases, the jet velocity decreases while the jet radius increases for the same compound layer fraction due to the increasing viscoelastic stresses. In addition, fewer jet drops are ejected as the jets become slower and broader with increasing polymer concentration, as viscoelastic stresses persist throughout the jet formation and thinning process. We further obtain a regime map delineating the conditions for jet drop ejection versus no jet drop ejection in bursting bubbles coated with a polymeric compound layer. Our results may provide new insights into the mechanisms of mass transport of organic materials in bubble-mediated aerosolization processes.
\end{abstract}



\section{Introduction}

In nature, countless bubbles are continuously formed through natural physical processes every second, such as wave breaking$\cite{deike2022mass}$, impact of raindrops$\cite{thoroddsen2018singular}$, and gas release from natural seeps$\cite{johansen2017time,johansen2020hydrocarbon}$. Bubbles are also utilized in a variety of industrial processes involving gas fluxing, such as in bioreactors$\cite{mcclure2016characterizing}$ and wastewater treatment$\cite{sakr2022critical}$. When these bubbles rise to the air-water interface due to buoyancy, they ultimately burst after the cap film ruptures. The subsequent collapse of the bubble cavity generates capillary waves that converge at the base of the bubble cavity, producing a Worthington jet, which can further disintegrate into smaller jet drops$\cite{deike2018dynamics}$. These drops could transport chemical (sea salts/toxins/microplastics)$\cite{lewis2004sea, textor2006analysis, veron2015ocean, olson2020harmful, deike2022mass, shaw2023ocean}$ and biological (bacteria/virus)$\cite{poulain2018biosurfactants,blanchard1989ejection,mcrae2021aerosol}$ substances into the atmosphere, impacting climate dynamics, earth system modeling, and public health$\cite{wilson2015marine,neel2022role,bourouiba2021fluid,bourouiba2021fluid2}$. Therefore, jetting dynamics from bursting bubbles play a vital role in controlling the mass transport across the air-liquid interface, and have received significant attention from researchers across disciplines$\cite{tammaro2021flowering, shaw2023ocean,veron2015ocean,chen2020bubble,zenit2018fluid,sakr2022critical,saththasivam2016overview,robinson2019rising,blanchard1989ejection,poulain2018biosurfactants}$.

\begin{table}[!ht]
    \centering
    \renewcommand{\arraystretch}{1.5} 
    \fontsize{10pt}{13pt}\selectfont 
   \begin{tabular}{
        >{\raggedright\arraybackslash}p{4cm}
        >{\raggedright\arraybackslash}p{2.2cm}
        >{\raggedright\arraybackslash}p{4cm}
        >{\raggedright\arraybackslash}p{3.0cm}
    }
        \toprule\toprule
        \textbf{Reference} & \textbf{Focus} & \textbf{Fluid Rheological model} & \textbf{Important Dimensionless Numbers} 
        \\
        \midrule
        Sanjay, Lohse, and Jalaal$\cite{sanjay2021bursting}$ & Numer. & Bulk: Viscoplastic Bingham model& \( 0 < \mathcal{J} < 64 \) 
        \\
        Rodr\'iguez-D\'iaz et al.$\cite{rodriguez2023bubble}$ & Exptl. & Bulk: Weakly viscoelastic aqueous PEO solutions & \( 10^{-7} < De < 10^{-3} \) 
        \\
        Cabalgante-Corralesa et al.$\cite{cabalgante2024effect}$ & {Exptl./Numer.} & Bulk: Viscoelastic Oldroyd-B model& \( 10^{-3} < De < 1 \), \( 10^{-3} < Ec < 10^{-1} \) 
        \\
        Dixit et al.$\cite{dixit2024viscoelastic}$ & Numer. & Bulk: Viscoelastic Oldroyd-B model & \( 10^{-4} < De < 10^{4} \), \( 10^{-4} < Ec < 10^{4} \) 
        \\
        Balasubramanian et al.$\cite{balasubramanian2024bursting}$ & Numer. & Bulk: Elastoviscoplastic model by Saramito$\cite{saramito2007new}$ & \( 10^{-3} < De < 30 \), \( 10^{-3} < \mathcal{J} < 10 \) 
        \\
        Current work & Exptl. & Coating compound: Weakly viscoelastic aqueous PEO solutions; Bulk: Newtonian& \( 10^{-4} < De < 10^{-2} \), \( 10^{-1} < Ec < 10 \), \( 0 < \psi_0 < 60\%  \) 
        \\
        \bottomrule\bottomrule
    \end{tabular}
    \caption{Summary of previous investigations for dynamics of bubble bursting jets in non-Newtonian fluids. The non-dimensional numbers are defined as follows: plastocapillary number ($\mathcal{J}$, the ratio between capillary and yield stresses), Deborah number ($De$; the ratio between the extensional relaxation time of the polymer solution to the inertio-capillary timescale), and elastocapillary number ($Ec$; the ratio between elastic and capillary stresses). $\psi_0$ is the volume fraction of the polymeric compound layer of bubbles.}
    \label{tbl:summary}
\end{table}

While most previous studies have focused primarily on clean bubbles, the jetting dynamics of contaminated bubbles have attracted considerable attention only recently. Rising bubbles can scavenge contaminants from biological or industrial origins$\cite{ji2022water,poulain2018biosurfactants,blanchard1989ejection,shaw2023ocean,tammaro2021flowering,robinson2019rising,constante2021dynamics,yang2024effect,ji2023secondary,decho2017microbial}$, e.g. surfactants, proteins, and biological gels, but their effects on the bubble bursting jets remain largely unexplored. These contaminants are known to modify the interfacial dynamics substantially by altering surface tension and creating surface tension gradients, i.e. Marangoni effects, as well as complicating the interfacial
rheology. Previous studies have shown the bursting of surfactant-laden bubbles generates fewer jet drops compared to surfactant-free cases due to the suppression of Marangoni stresses$\cite{constante2021dynamics,pierre2022influence,pico2024surfactant}$. For bubble bursting at a protein-laden bubble interface, surface elasticity significantly alters the dynamics of cavity collapsing, reducing jet velocity while increasing jet radius$\cite{ji2023secondary,yang2024effect}$. Specifically, biochemical contaminants, such as microbial extracellular polymeric substances (EPSs), may form a viscoelastic layer at the bubble surface due to the three-dimensional network of organic exopolymers$\cite{decho2017microbial,mrokowska2019viscoelastic}$. Furthermore, EPSs have been identified as one of the key components in marine aerosols ejected by bubble bursting, contributing to cloud condensation nuclei and impacting global radiation$\cite{bigg2007sources, bigg2008composition, leck2008comparison, kuznetsova2005characterization}$. Consequently, understanding the influence of non-Newtonian rheology on bubble bursting behavior is essential for advancing our understanding of marine biology and environmental science.

We summarize the recent studies for the effect of non-Newtonian fluid rheology on bubble bursting jets in Table $\ref{tbl:summary}$. For bare bubble bursting jets in a non-Newtonian fluid, previous numerical investigations have discussed the effect of a viscoplastic$\cite{sanjay2021bursting}$, viscoelastic$\cite{dixit2024viscoelastic}$, or elastoviscoplastic$\cite{balasubramanian2024bursting}$ medium on the bubble cavity collapse and jet ejection. Specifically, multiple non-dimensional numbers, including the plastocapillary number ($\mathcal{J}$, the ratio between yield stress and capillary pressure), the Deborah number ($De$, the ratio between the extensional relaxation time of the polymer solution to the inertio-capillary timescale), and the elastocapillary number ($Ec$, the ratio between elastic and capillary stresses), are used to describe the bulk non-Newtonian rheology. Meanwhile, to the best of our knowledge, systematic experimental investigations in this area remain significantly limited. Using low-molecular-weight polyethylene oxide solutions, Rodr\'iguez-D\'iaz et al.$\cite{rodriguez2023bubble}$ observed that weak viscoelasticity suppresses the ejection of jet drops. A follow-up recent experimental study by Cabalgante et al.$\cite{cabalgante2024effect}$ showed that the polymer viscosity has the largest effect on the jet velocity while the polymer relaxation time affects where a jet drop is emitted or not. However, previous studies have not explored a widely encountered scenario in nature and industry: compound bubbles coated by a viscoelastic layer similar to EPS. Compared with bare bubble bursting in a non-Newtonian bulk medium, such compound bubbles may result in distinct jetting dynamics, which are more pertinent to the transport and fate of EPS in real oceanic environments, therefore requiring particular attention.

Here, we experimentally investigate the bursting dynamics of compound bubbles coated with a viscoelastic layer, focusing on their jetting dynamics and implications for aerosol generation. We construct our paper as follows: The experimental setup and the rheological characterization for the working fluids are described in Section 2. In Section 3, we analyze the influence of the polymeric compound layer on cavity collapse and the resulting bubble bursting jets, including their corresponding top jet drops. These observations are made across a range of polymer concentrations and compound layer volume fractions, with each parameter quantitatively assessed. Additionally, we discuss the effects of the compound layer on jet dynamics, particularly considering the non-Newtonian rheology. A regime map is also provided for the first time, illustrating the jet drop/no-jet drop regimes based on variations in polymer concentration and compound layer fraction. Finally, we conclude our discoveries and implications of this study in Section 4.

Here, we experimentally investigate the bursting dynamics of compound bubbles coated with a viscoelastic layer, focusing on their jetting dynamics and implications for aerosol generation. We construct our paper as follows: The experimental setup and the rheological characterization for the working fluids are described in Section 2. In Section 3, we analyze the influence of the polymeric compound layer on cavity collapse and the resulting bubble bursting jets, including their corresponding top jet drops. These observations are made across a range of polymer concentrations and compound layer volume fractions, with each parameter quantitatively assessed. Additionally, we discuss the effects of the compound layer on jet dynamics, particularly considering the non-Newtonian rheology. A regime map is also provided for the first time, illustrating the jet drop/no-jet drop regimes based on variations in polymer concentration and compound layer fraction. Finally, we conclude our discoveries and implications of this study in Section 4.

\section{Experimental Methodology}

\begin{figure*}[ht!]
    \centering
    \includegraphics[width=0.7\linewidth]{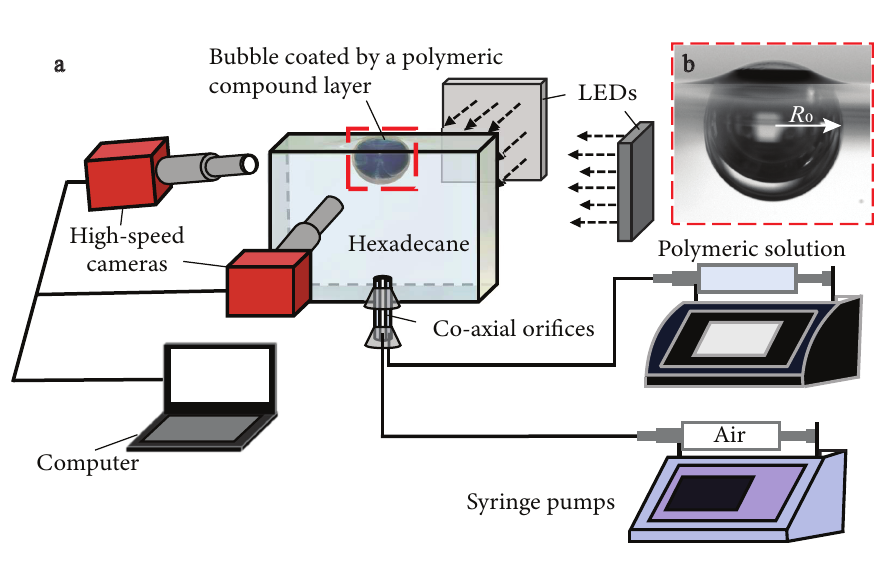}
    \caption{(a) Experimental setup for high-speed imaging for the jetting dynamics of bubbles with a viscoelastic compound interface. (b) Zoomed-in image of a typical compound bubble coated by a polymeric layer. $R_{0}$ is the compound bubble radius.}
    \label{fig:expSetup}
\end{figure*}

\textbf{Experimental setup.} Figure \ref{fig:expSetup} shows the schematics of our experimental setup. A co-axial orifice system was used to generate compound bubbles in a controlled way$\cite{ji2021oil,ji2021oil2}$. Two syringe pumps (PHD ULTRA and 11 Pico Plus Elite, Harvard Apparatus) were connected to the outer and inner needles of the orifice system to infuse the polymer solution and air at controlled flow rates. The bubble was released into an acrylic container large enough to minimize wall effects on bubble bursting dynamics. A slightly convex meniscus was maintained at the top of the container to keep the coated bubble at the center. Two high-speed cameras (FASTCAM Mini AX200, Photron) were used to synchronously capture the temporal evolution of the cavity collapse and jetting dynamics above and below the free surface, respectively. Both cameras operated at a frame rate of 6400 fps with an image resolution of 5.6-14.3~\text{µm/px}.  All experiments were conducted with a compound bubble radius of $R_{0}$= 1.48 $\pm$ 0.11 mm as shown in Fig. \ref{fig:expSetup}. The volume fraction of the compound layer is defined as $\psi_0= 3V_0/ (4\pi R_{0}^3$), where we calculate the volume of the compound layer $V_0$ by image analysis right before jet formation$\cite{yang2023enhanced}$.

 \begin{figure*}[!h]
    \centering
    \includegraphics[width=0.65\linewidth]{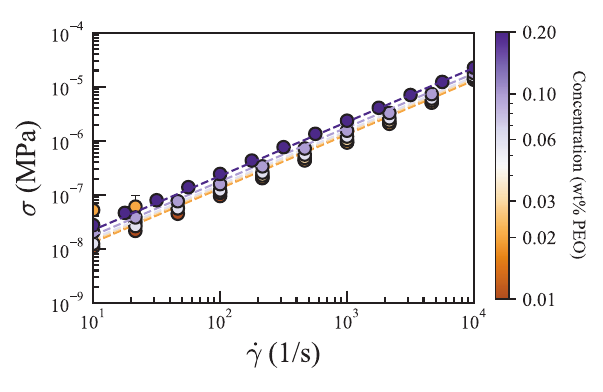}
    \caption{ Shear stress $\sigma$ as a function of the shear rate $\dot{\gamma}$ for PEO solutions of different concentrations.}
    \label{fig:viscositycurves}
\end{figure*}

\noindent \textbf{Materials.} We used aqueous solutions of polyethylene oxide (PEO) (Sigma-Aldrich, molecular weight of $6 \times 10^{5}$ g/mol) as the viscoelastic compound layer. The solutions were prepared by dissolving the polymers in deionized water (Smart2Pure 3 UV/UF, ThermoFisher Scientific, 18.2 $\text{M}\Omega\cdot\text{cm}$ at $20^\circ\text{C}$) on a magnetic stirrer for 24 hours, at no heat and low stirring rates to minimize thermal and mechanical degradation$\cite{rodriguez2023bubble}$. The PEO solutions are chosen to be sufficiently dilute to be described as Boger fluids without significant shear thinning$\cite{rodriguez2023bubble,james2009boger}$. All PEO concentrations used in this study are below the critical overlap concentration estimated as 2.44 wt\%, which defines the upper limit below which a polymer solution is considered dilute. Above this threshold, polymer chains start to enter a semi-dilute regime to overlap, interact, and may thereafter form entangled networks$\cite{tirtaatmadja2006drop, del2017relaxation, james2009boger}$. As shown in Fig.~\ref{fig:viscositycurves} b), we performed rheological measurements (TA Instruments DHR-3 with a 40mm diameter and $1^\circ$ cone plate) and confirmed that there are no considerable shear-thinning effects. To further characterize the apparent extensional relaxation time, we employ the semi-empirical curve fit proposed by Rodríguez-Díaz et al. as $\lambda_r = 2.707 \times 10^{-7} c^{1.733}$, where $\lambda_r$ represents the apparent extensional relaxation time in ms and $c$ denotes the polymer concentration in parts per million (ppm) for PEO$\cite{rodriguez2023bubble, rubio2020experimental}$.
Hexadecane (Sigma-Aldrich, Reagent Plus, 99\%, density $\rho_b=$ 773 kg/m$^3$, dynamic viscosity $\mu_b=$ 3.45 mPa s) was used as the Newtonian bulk liquid phase. The interfacial tensions were measured using the pendant drop method and analyzed using the open-source software Opendrop$\cite{huang2021opendrop}$. All the fluid properties are summarized in Table \ref{tbl:materials}.

\begin{table*}[h!]
  \centering
  \resizebox{0.8\textwidth}{!}{ 
    \begin{tabular}{llllll}
      \hline
      PEO concentration, wt\% & $\eta_{t}$, mPa s & $\eta_{p}$, mPa s & $\lambda_r$, ms & $\gamma_{ac}$, mN/m & $\gamma_{cb}$, mN/m \\
      \hline
          0.01 & 1.35 $\pm$ 0.02 &  0.35 $\pm$ 0.02 & 0.001 & 59.7 $\pm$ 0.5 & 29.8 $\pm$ 0.6 \\
      0.02 & 1.32 $\pm$ 0.02 & 0.32 $\pm$ 0.02 & 0.003 & 61.3 $\pm$ 0.5 & 30.0 $\pm$ 0.1 \\
      0.03 & 1.44 $\pm$ 0.03 & 0.44 $\pm$ 0.03 & 0.005 & 56.9 $\pm$ 0.3 & 28.2 $\pm$ 0.1 \\
      0.06 & 1.54 $\pm$ 0.08 & 0.54 $\pm$ 0.08 & 0.018  & 61.8 $\pm$ 0.4 & 29.7 $\pm$ 0.2 \\
      0.1 & 1.77 $\pm$ 0.02 & 0.77 $\pm$ 0.02 &  0.043  & 61.6 $\pm$ 0.4 & 26.9 $\pm$ 0.6 \\
      0.2 & 2.24 $\pm$ 0.03 & 1.24 $\pm$ 0.03 &  0.142  & 61.1 $\pm$ 0.4 & 29.4 $\pm$ 0.3 \\
    \hline
    \end{tabular}
  }
  \caption{Material properties of the working fluids. Here, $\lambda_r$ is the extensional relaxation time, $\gamma$ is the interfacial tension, and the subscripts $a$, $b$ and $c$ represent air, bulk, and coating compound phases, respectively. }
  \label{tbl:materials}
\end{table*}

\noindent {\textbf{Dimensionless numbers.} Based on previous numerical investigations for bubble bursting in a bulk non-Newtonian fluid modeled with Oldroyd-B
viscoelastic behavior$\cite{dixit2024viscoelastic,cabalgante2024effect}$, we characterize the non-Newtonian rheological effects with the following dimensionless numbers related to the relaxation time $\lambda_r$ and the elastic modulus $G= {\eta_p}/{\lambda_r}$ of the polymeric solution$\cite{dixit2024viscoelastic}$, respectively. Here, $\eta_p$ represents the polymer viscosity, calculated as $\eta_p = \eta_{t} - \eta_s$, where $\eta_{t}$ is the total viscosity of the polymeric solution and $\eta_s$ is the solvent viscosity. Given that the bubble cavity collapse dynamics occur on the order of the inertio-capillary time scale ${t_c}=\sqrt{\rho_b R_0^3/\gamma_e}$, we use the Deborah number ${De} ={\lambda_r}/{t_c}$ to describe the effect of polymer relaxation on the bursting dynamics. Here, $\rho_b\ \mathrm{and}\ R_0$ represent bulk liquid density and bubble radius, respectively, and the effective surface tension $\gamma_e= \gamma_{ac}+ \gamma_{cb}$ is calculated as the sum of air-compound layer ($\gamma_{ac}$) and compound layer-bulk ($\gamma_{cb}$) interfacial tensions. Meanwhile, we use the elastocapillary number $Ec = {G R_0}/{ \gamma_e}$ to characterize the ratio between the elastic and capillary stresses. In addition, the Ohnesorge number that compares the inertial–capillary to inertial–viscous timescales is also considered. Specifically, we introduce the polymeric and solvent Ohnesorge numbers $Oh_{p}=\eta_p/\sqrt{\rho_c\gamma_{e}R_0}$ and $Oh_{s}=\eta_s/\sqrt{\rho_c\gamma_{e}R_0}$ to describe the effects of the polymer and solvent viscosity, respectively, where $\rho_c$ is the compound layer density. Table \ref{tbl:nondim} lists the calculated dimensionless number for all working fluids. We note that the gravity effect is considered negligible in current experiments given a small Bond number $Bo=\rho_b g R_{\text{0}}^2 / \gamma_e$ (ratio between gravity and capillary effects) of $\approx$ 0.19.

\begin{table*}[h!]
  \centering
  \resizebox{0.8\textwidth}{!}{ 
    \begin{tabular}{llllll}
      \hline
      PEO concentration, wt\% & $De$ & $Ec$ & $Oh_{p}$ & $Oh_{s}$ & $Oh_{t}$ \\
      \hline
        
        $0.01$ & $1.50 \times 10^{-4}$ & $7.39$ & $9.70 \times 10^{-4}$ & $2.75 \times 10^{-3}$ & $3.72 \times 10^{-3}$ \\
        
        $0.02$ & $5.02 \times 10^{-4}$ & $1.95$  & $8.59 \times 10^{-4}$ & $2.71 \times 10^{-3}$ & $3.57 \times 10^{-3}$ \\
        
        $0.03$ & $9.79 \times 10^{-4}$ & $1.44$ & $1.24 \times 10^{-3}$ & $2.81 \times 10^{-3}$ & $4.05 \times 10^{-3}$ \\
        
        $0.06$ & $3.37 \times 10^{-3}$ & $4.92 \times 10^{-1}$ & $1.46 \times 10^{-3}$ & $2.71 \times 10^{-3}$ & $4.17 \times 10^{-3}$ \\
        
       $0.1$ & $8.04 \times 10^{-3}$ & $3.02 \times 10^{-1}$ & $2.13 \times 10^{-3}$ & $2.76 \times 10^{-3}$ & $4.89 \times 10^{-3}$ \\
        
       $0.2$ & $2.70 \times 10^{-2}$ & $1.43 \times 10^{-1}$ & $3.39 \times 10^{-3}$ & $2.73 \times 10^{-3}$ & $6.12 \times 10^{-3}$ \\
      \hline
    \end{tabular}
  }
  \caption{Ranges of non-dimensional numbers in current experiments.}
  \label{tbl:nondim}
\end{table*}

\section{Results and discussion}
\vspace{\baselineskip}

\subsection{\textbf{\small Cavity collapse and capillary wave propagation}}

Figure \ref{fig:psi100and2000} shows the cavity shape evolution from the cap film breakage (t=0 ms) to the interface reversal, including bare bubble and compound bubble bursting in pure hexadecane. When the bubble cap ruptures, the cavity collapse generates capillary waves that propagate along the air-compound layer interface, contributing to the jet formation ultimately. 
The capillary waves are progressively damped as they travel down the cavity bottom, substantially affecting the final jet formation. 
During cavity collapse of a compound bubble, the polymeric compound layer retracts toward the cavity bottom, forming a bulb structure. This retraction behavior is associated with the unfavorable wetting characteristics of the polymeric solution, as verified by the negative spreading coefficient $S= \gamma_{ab}-\gamma_{ac}-\gamma_{cb}<0$ for all our cases$\cite{de2004capillarity}$. Here, $\gamma_{ab}$ is the surface tension of the bulk hexadecane. In addition, at the end of the cavity collapse, we start to observe the entrainment of polymer threads from the collapsing cavity right around jet birth for high PEO concentration and compound layer volume fraction (inset of Figure \ref{fig:psi100and2000} (c)). The large surface compression during cavity collapse results in the enrichment of the PEO molecules absorbed onto the cavity surface, which get entrained into the compound layer by the extensional flow produced by bubble bursting, similar to the protein fragments shedding from a compressed protein-adsorbed bubble surface reported previously$\cite{ji2023secondary}$.

A key parameter to characterize the capillary waves is the wavelength $L$ measured between the last two consecutive wave troughs of the capillary wave train $\cite{krishnan2017scaling,gordillo2019capillary,yang2023enhanced,yang2024effect}$. For compound bubbles, capillary waves separate onto the air-compound and compound-bulk interfaces, causing a decrease in $L$ with $\psi_0$ compared with bare bubble bursting, as shown in Fig. \ref{fig:psi100and2000} at around 0.25$t_c$. Such a decrease of the characteristic wavelength increases the damping rate as discussed by the previous work \cite{lamb1924hydrodynamics,yang2023enhanced}, thus results in a narrower jet base. To further understand the influence of the polymeric compound layer on the cavity collapse, we also examine the degree to which the cavity forms a cone-like shape right before the jet formation. A geometric dependence has been established in a previous work$\cite{gordillo2023theory}$ between the kinematic properties of the jet and the semi-angle of the cavity cone formed when the capillary waves converge at the cavity nadir: the jet velocity increases and the jet radius decreases with the decreased cone angle. Figure \ref{fig:psi100and2000} (b) and (c) show that the cone angle (defined as $2\beta$) for the higher $\psi_0$ is smaller than that of the lower $\psi_0$, indicating a progression towards the singular limit of cavity collapse in the former case. We will elaborate in subsequent sections on how the cavity collapse influences the jetting dynamics. Notably, Figure \ref{fig:collapse} shows that the cavity collapse time ($t_{cc}$) remains unchanged as the polymer concentration varies from the lowest (\( c = 0.01 \, \text{wt}\% \)) to the highest (\( c = 0.2 \, \text{wt}\% \)) in the experiments and is also unaffected by the compound layer volume fraction. This is similar to bare bubble bursting in a weakly viscoelastic medium where the cavity collapse time also does not change with polymer concentration$\cite{rodriguez2023bubble,dixit2024viscoelastic,cabalgante2024effect}$. While PEO molecules are excepted to adsorb onto the interface, we infer that Marangoni stresses or interfacial rheology do not significantly influence the cavity collapsing and jetting dynamics in current experiments with \(10^{-4} < De < 10^{-2} \), \( 10^{-1} < Ec < 10 \), \( 0 < \psi_0 < 60\% \), since no changes are observed in the timing of cavity collapse at any $c$ and $\psi_0$.

\begin{figure*}[!ht]
    \centering    
    \includegraphics[width=1\linewidth]{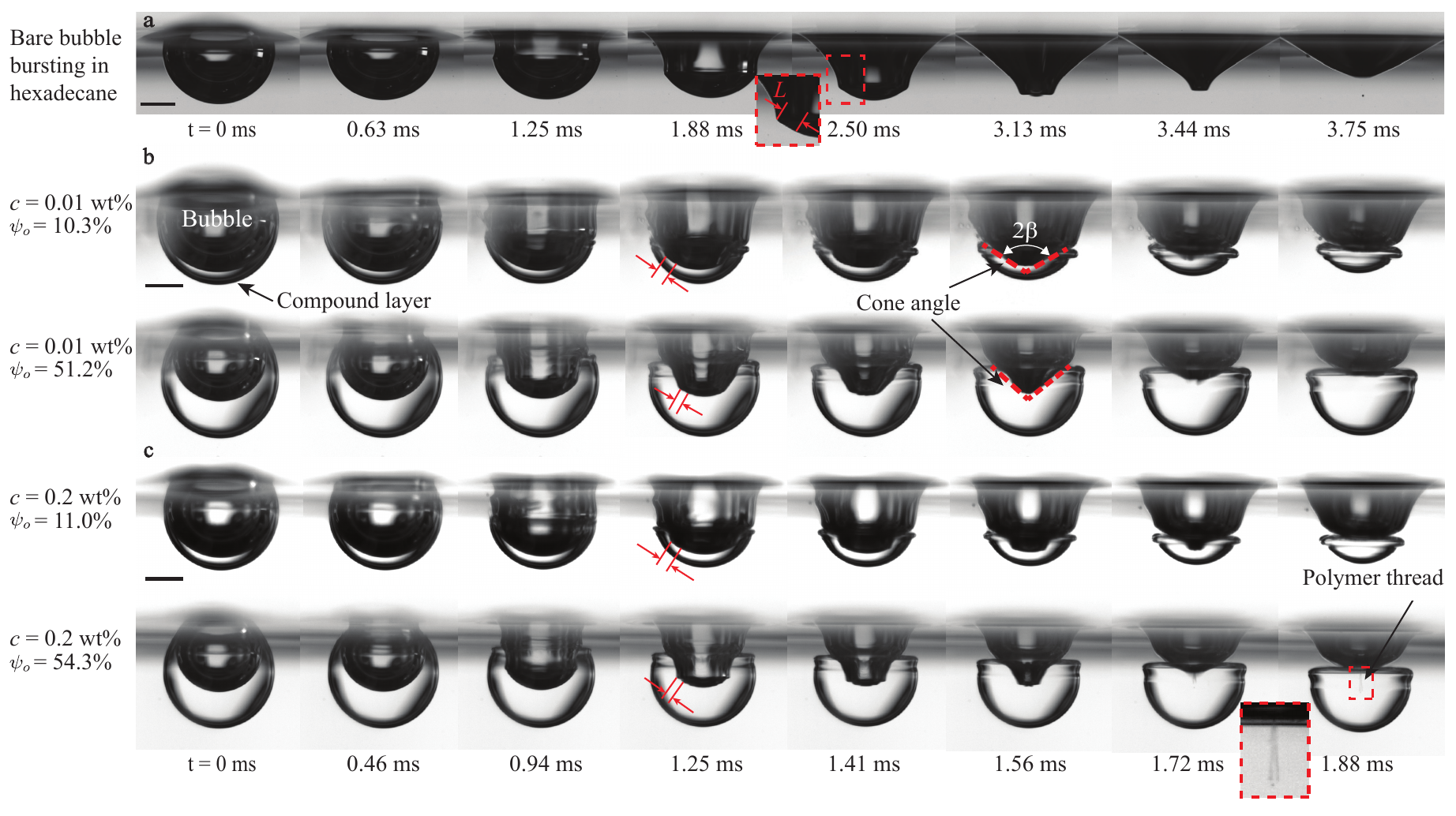}
    \caption{High-speed imaging of bubble cavity collapse: (a) bare bubble in hexadecane, (b) compound bubble coated by a compound layer with $c$=0.01 wt\% at $\psi_0$=10.3\% and 51.2\%, and (c) compound bubble coated by a compound layer with $c$=0.2 wt\% at $\psi_0=$11.0\% and 54.3\%. The wavelength $L$ was measured between the last two consecutive wave troughs of the capillary wave train. The red dashed lines denote the cone angle of the cavity geometry 2$\beta$ right before jet formation. The inset in (c) shows a zoom-in view of the polymer thread entrained by bubble bursting flows. All scale bars represent 1 mm. See also Supplementary Videos 1 and 2 corresponding to (c).}
    \label{fig:psi100and2000}
\end{figure*}

\begin{figure*}[!h]
    \centering
    \includegraphics[width=0.6\linewidth]{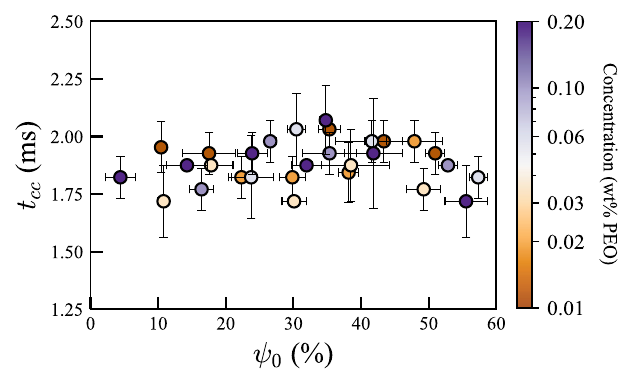}
    \caption{Cavity collapse time $t_{cc}$ as a function of $\psi_0$ for bursting bubbles coated by PEO solutions with concentrations of 0.01 - 0.2 wt\%. The collapse time across all cases remains nearly constant, with a value of 1.88 $\pm$ 0.11 ms.}
    \label{fig:collapse}
\end{figure*}

\subsection{\textbf{\small Jetting dynamics}}

Following the cavity collapse and focusing of capillary waves at the cavity bottom, jet ejection occurs with the reversal of the bottom curvature as shown in Figures \ref{fig:c-constant} and \ref{fig:psi-constant} from systematically controlled experiments varying with both the volume fraction $\psi_0$ and polymer concentration $c$. The rising Worthington jets primarily consist of polymer solutions, which undergo substantial extensional deformation during their formation and ascent. Due to the Rayleigh-Plateau instability, the jets break into drops, each connected by a viscoelastic filament that ultimately ruptures, releasing the drops. As $\psi_0$ increases while the polymer concentration $c$ maintains constant, a comparison between Figures \ref{fig:c-constant}(a) and (b) reveals that the jet grows thinner and faster, producing drops noticeably smaller and of a larger number. Additionally, a characteristic ``beads-on-a-string" structure emerges in the cases with drop formation. During the jet rise, the viscoelastic filament undergoes tensile stretching while connecting drops for a prolonged time until its eventual breakup, which leads to the ejection of jet drops. The number of drops increases with $\psi_0$ for a given $c$. As the polymer concentration $c$ increases for a specific $\psi_0$, as shown in Fig.~\ref{fig:psi-constant} (a) and (b), the jet becomes thicker and weaker, showing a stronger inhibition on the jet ejection with increasing viscoelastic effect. Fewer or even no drops are produced due to the widening of the jet shape. In all experimental cases, the viscoelastic compound layer fluid is consistently entrained into the jet. In cases where drop ejection occurs, jet drops are predominantly composed of the compound layer fluid and coated with a thin layer of the bulk Newtonian fluid. Furthermore, at higher polymer concentrations ($c \geq 0.2$ wt\%), jet drop formation is completely suppressed, and only a rising jet containing the compound layer fluid is observed.

\begin{figure*}[!ht]
    \centering
    \includegraphics[width=0.8\linewidth]{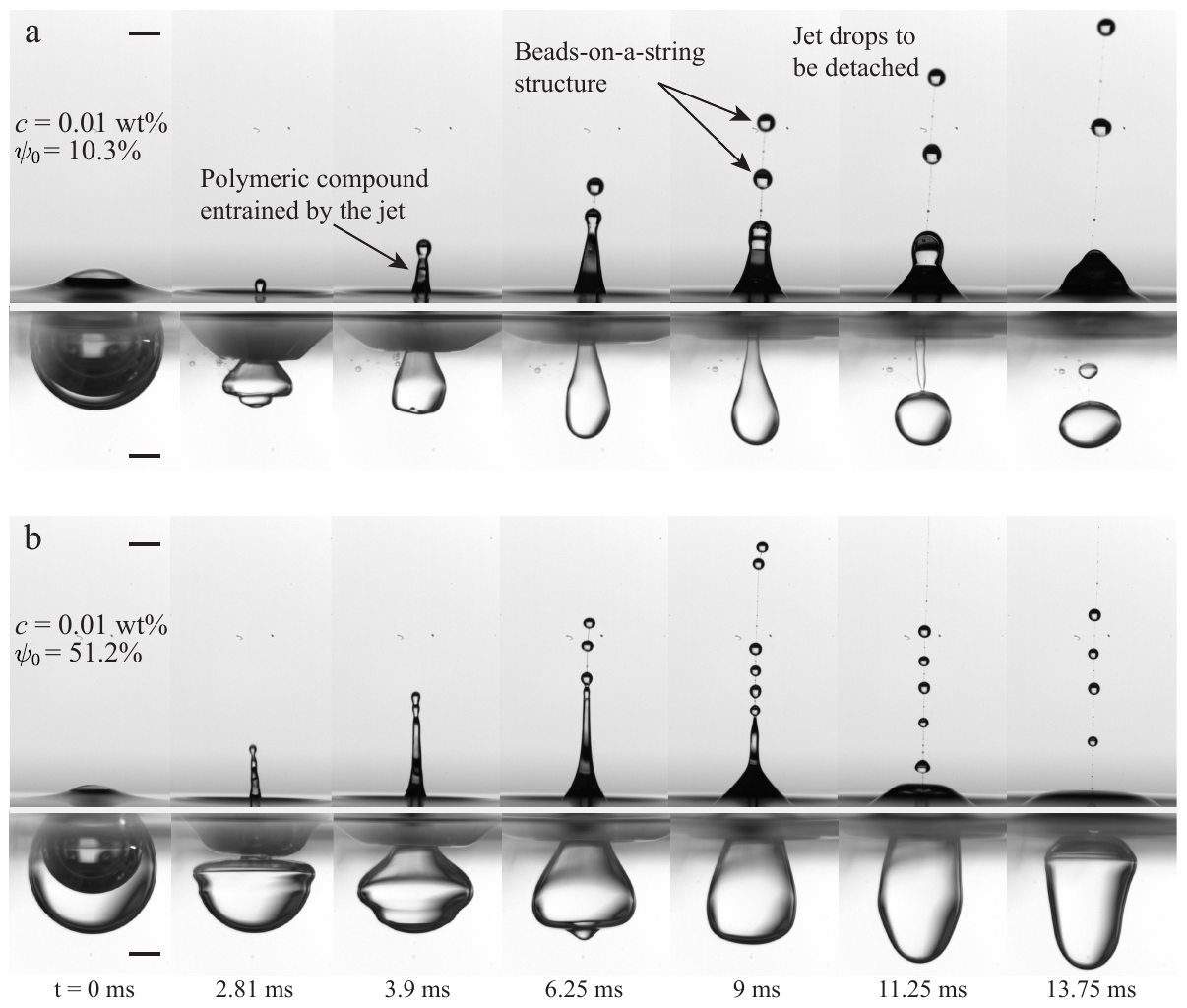}
    \caption{Side view of a bursting bubble coated by a compound layer at a PEO concentration of 0.01 wt\% for (a) $\psi_0 = 10.3\%$ and (b) $\psi_0 = 51.2\%$. During the rising of the jet, the end-pinching instability causes it to break up into drops and form a beads-on-a-string structure. All scale bars represent 1 mm. See also Supplementary Videos 3 and 4 corresponding to (a).}
    \label{fig:c-constant}
\end{figure*}

\begin{figure*}[!ht]
    \centering
    \includegraphics[width=0.85\linewidth]{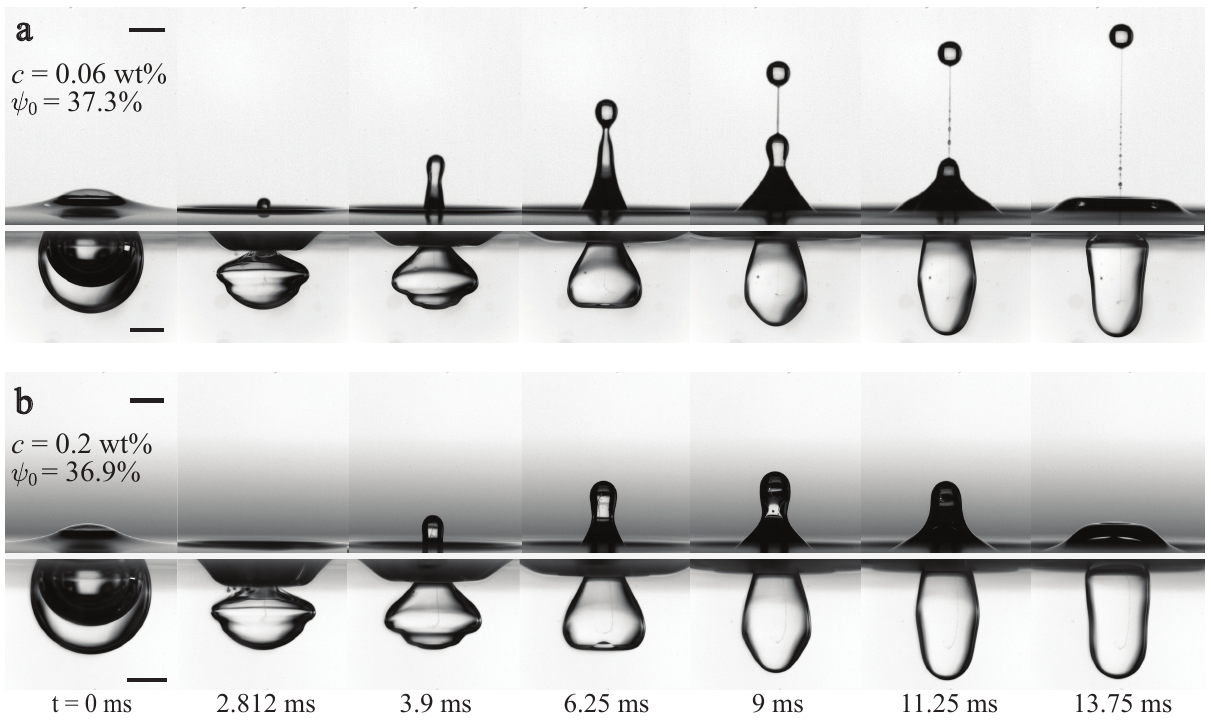}
    \caption{Side view of a bursting bubble coated by a compound layer at a PEO concentration (a) $c$= 0.06 wt\% and (b) $c$= 0.2 wt\% for $\psi_0 \approx 37\%$. As $c$ increases while maintaining a nearly constant $\psi_0$, we observe a transition in compound bubble bursting behavior from generating jet drops to producing no jet drops. All scale bar represents 1 mm.}
    \label{fig:psi-constant}
\end{figure*}

We further measure the nondimensionalized jet velocity $v_j/v_{ce}$ and nondimensionalized jet radius $r_j/R_0$ as a function of compound layer volume fraction $\psi_0$ at different PEO concentrations, as shown in  Figure \ref{fig:vj}. Here, jet velocity $v_j$ is non-dimensionalized by the capillary velocity $v_{ce}= \sqrt{\gamma_e / (\rho_b R_{0})}$, and jet radius $r_j$ is non-dimensionalized by the compound bubble radius $R_0$. We measured the jet velocity and radius, both when the jet tip crosses the undisturbed bulk free surface level. 
Compared to the bare bubble bursting case with a similar bubble radius, the dimensionless jet velocities produced by compound bubble bursting cases are smaller, while the dimensionless jet radii are larger. 
This observation signifies that the viscoelastic compound layer on a bubble suppresses the jet ejection compared to the bare bubble case, highlighting the necessity of understanding the role of viscoelastic effects in bubble-bursting jet formation.

When the  compound layer volume fraction $\psi_0$ increases for a constant polymer concentration $c$,
we observe that \(v_j\) increases  while \(r_j\) decreases, until they plateau at \(\psi_0 \geq 30\%\). The increase in jet velocity as a function of $\psi_0$ for the same $c$ is attributed to the enhancement of the jet due to the damping of short-wavelength precursor capillary waves during cavity collapse. Larger $\psi_0$ results in smaller characteristic wavelength as a thicker compound layer decreases the wavelength $L$ more significantly due to earlier and stronger wave separation $\cite{yang2023enhanced}$, leading to less short-wavelength perturbation for the focusing of the capillary waves at the cavity nadir which allows the formation of a faster and thinner jet. Additionally, capillary wave focusing at a higher \(\psi_0\) produces a cavity with a lower cone angle, $2\beta$, for the same \(c\), as showcased in Figure \ref{fig:psi100and2000}. Such smaller cone angle has been shown to favor the production of narrower jets with a faster speed in previous theoretical and simulation investigation$\cite{gordillo2023theory}$, which is also consistent with the enhancement of jetting observed in our experiments. 

Meanwhile, $v_j/v_{ce}$ decreases and $r_j/R_0$ increases significantly with PEO concentration, while $Oh_{t}$ remains relatively constant. As PEO concentration increases from 0.01 wt\% to 0.2 wt\%, $De$ increases by two orders of magnitude from $1.50 \times 10^{-4}$ to $2.70\times 10^{-2}$. The substantial increase in polymer relaxation time suggests enhanced viscoelastic effects. The stronger viscoelastic effect of the compound layer leads to a thicker, slower, and wider jet with a reduced final height, which is consistent with similar studies of bubble bursting in viscoelastic liquids$\cite{dixit2024viscoelastic,rodriguez2023bubble,cabalgante2024effect}$. 
 In essence, at the moment of jet formation in the polymeric compound layer, the axial strain rate and viscoelastic stresses at the jet base increase due to polymer stretching caused by the extension flow. As $De$ increases, the extensional thickening is strengthened by increased elasticity due to higher polymer concentrations.

\begin{figure*}[!ht]
    \centering
    \includegraphics[width=1\linewidth]{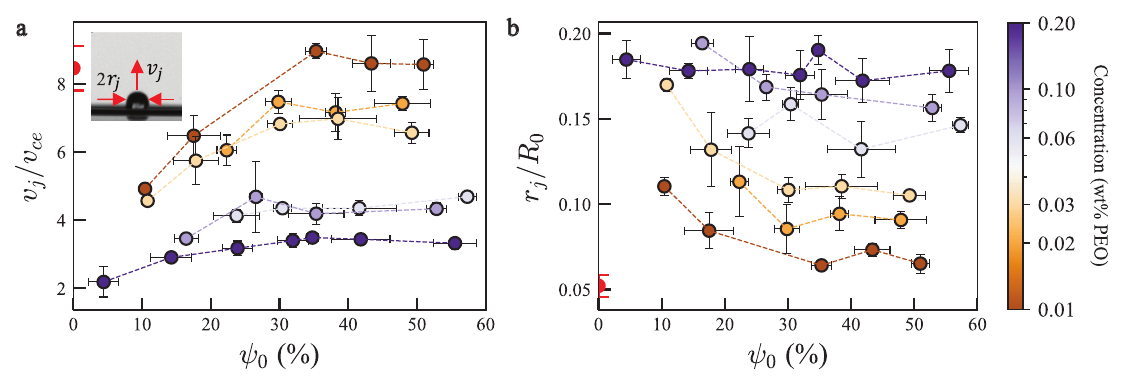}
    \caption{(a) Non-dimensionalized jet (a) velocity and (b) radius as functions of compound layer volume fraction for bubble bursting with PEO concentrations of 0.01 - 0.2 wt\%. Red markers and error bars denote the case of bare bubble bursting in hexadecane, with a bubble radius similar to that of a compound bubble.}
    \label{fig:vj}
\end{figure*}

\subsection{Velocity and radius of top jet drops}
At the end stages of the jetting, we observe that drops pinch off from the end of the jet due to Rayleigh-Plateau instability. The pinched-off drops form filaments between them and the main jet due to the transition from an inertio-capillary regime to the elastocapillary regime$\cite{sen2022elastocapillary, cabalgante2024effect}$. The elastocapillary thinning of these filaments, driven by the tensile forces exerted by the ejecting drops, gives rise to the characteristic "bead-on-a-string" structure. This filament-thinning is primarily governed by the interplay of surface tension and elasticity, persisting until the filament breaks and releases the drops.  In Fig.~\ref{fig:vd}, we show the dimensionless velocity $v_d/v_{ce}$ and radius $r_d/R_0$ of the top drop as a function of $\psi_0$.
The velocity and radius of the top jet drop are measured when the filament thins to a critical threshold ($\sim$ 30 $\mu$m).
The drop velocity increases and radius decreases for increasing $\psi_o$ and decreasing $c$, a consistent trend with that of the jet velocity and radius. 
Additionally, for the lowest polymer concentrations $c$, the velocity of the drop $v_d$ almost matches the jet velocity $v_j$, indicating minor influence from polymer filament stretching as $De \sim \mathcal{O}(10^{-4})$. However, for higher concentrations, a noticeable difference between $v_d$ and $v_j$ emerges. The difference between $v_j/v_{ce}$ to $v_d/v_{ce}$ increases from 2.2 to 3.8 in average as $c$ increases from 0.01 wt\% to 0.1 wt\%, demonstrate the significant viscoelastic effect with increasing $De$.  At the initial stage of drop formation, the velocity of the top jet drop matches the velocity of its jet. However, the drops decelerate from the initial velocity over time due to the pulling force from the thinning filament. As $De$ increases, a larger pulling force is exerted by the filament, which ultimately accounts for the significant difference between \(v_d\) and \(v_j\) at higher \(c\). This can be evidenced by comparing Figs. \ref{fig:c-constant} and \ref{fig:psi-constant}(a), as the viscous filament for $c=0.06$ wt\% maintains thicker and relaxes slower compared to those for $c=0.01$ wt\%, stretching on the ejected drops.  The radius of the ejected drops $r_d$, which are determined by the breakup dynamics of the jet and influenced by the jet rheological properties, similarly increases significantly for cases with higher PEO concentration and $De$.

\begin{figure*}[!ht]
    \centering
    \includegraphics[width=1\linewidth]{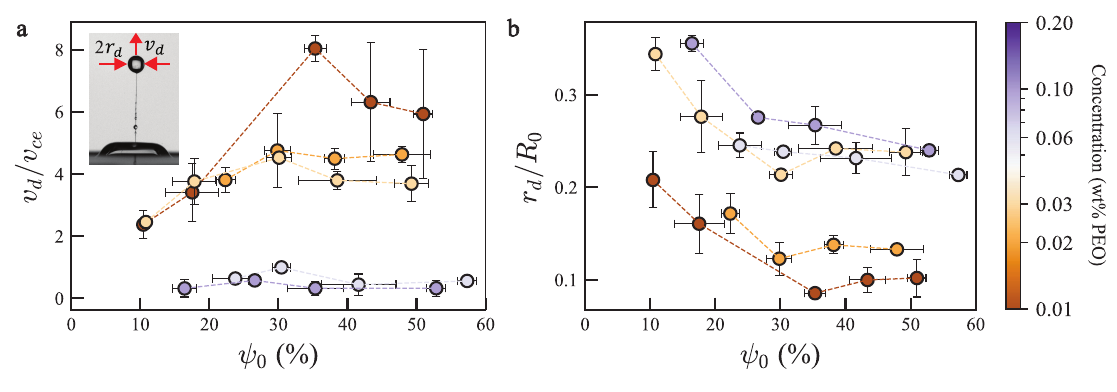}
    \caption{Non-dimensionalized top jet drop (a) velocity and (b) radius as a function of compound layer volume fraction with PEO concentrations of 0.01 - 0.1 wt\% under which jet drops form.}
    \label{fig:vd}
\end{figure*}

\subsection{Numbers of the jet drops}

We observe a systematic dependence of the number of jet drops ($N_d$) produced on $\psi_0$ for different polymer concentrations, as illustrated in Figure $\ref{fig:no of drops}$. Here, we account for all drops generated during bubble bursting, regardless of whether they detach from the thinning filament. The most significant trend is the increase of $N_d$ with $\psi_0$ for $c < 0.06$ wt$\%$. This is due to the effect of jet enhancement with increasing $\psi_0$ that results in higher $v_j$, for which thinner and more slender jet permits the formation of more drops. For $c=$ 0.06 or 0.1 wt$\%$, $N_d=1$ stays constant. A transition from a drop-producing regime to a no-jet-drop regime is observed as $c$ continues to increase, with drop ejection ceasing entirely at $c$ = 0.2 wt\% for all $\psi_0$ values. 
Additionally, we observe an overall decrease of $N_d$ with $c$. In cases with $c < 0.06$ wt$\%$, multiple drops form, with the initial drops ejected upward and completely detaching after filament thinning, while the remaining drops, lacking sufficient velocity to overcome the filament's retraction force, are pulled back into the bulk fluid. For cases where $0.06$ wt\% $\leq c\leq$ 0.1 wt\%, only a single drop is ejected, and it is eventually pulled back into the bulk for $c=0.1$ wt\%.  The decrease of $N_d$ is attributed to the increased viscoelastic effect. As previously discussed, larger $De$ results in a thicker filament due to a higher amount of polymer solution being entrained by the rising jet, leading to larger drop radii. With lower and thicker jets, fewer drops can be generated from the Rayleigh-Plateau mechanism. 

\begin{figure*}[!ht]
    \centering
    \includegraphics[width=0.6\linewidth]{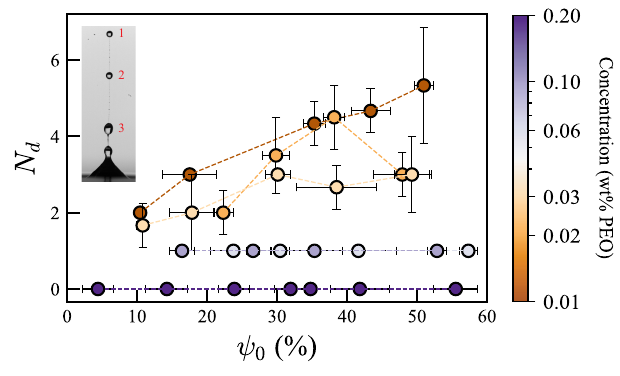}
    \caption{Number of jet drops, $N_d$, as a function of compound layer volume fractions with different PEO concentrations of 0.01 - 0.2 wt\%. The inset illustrates an example featuring three jet drops.}
    \label{fig:no of drops}
\end{figure*}

\subsection{Regime map for bubble bursting jet drops}

Based on the above results, we demonstrate a regime map of whether jet drops will be produced, as shown in Fig. \ref{fig:phase}.  
When comparing our compound bubble bursting cases with the previously reported $Oh$-$Bo$ regime map for jet drop production in bare bubble bursting$\cite{walls2015jet}$ (Fig. \ref{fig:phase} (Right)), current experiments fall within the predicted Newtonian jet drop region, including the no-jet-drop cases at the highest polymer concentration of \(c = \)0.2 \(\text{wt}\%\). The difference highlights that the effect of a viscoelastic compound layer could profoundly modify the jetting dynamics. We also show the regime map for jet drop production regarding the important dimensionless parameters in our experiments, $De$ and $\psi_0$ (Fig. \ref{fig:phase} (Left)). Notably, for the first time, we show that compound bubble bursting transitions into a no-jet-drop regime for \(De \gtrsim 10^{-2}\), where the ejected jet no longer produces jet drops. This is due to the relaxation time being sufficiently high to induce substantial polymeric stresses, leading to a corresponding increase in extensional viscosity. As a result, more of the polymeric coating is entrained, ultimately slowing down the jet and suppressing drop ejection in the short period of jet rising before it falls back to the pool. We note that \(\psi_0\) has negligible influence on the transition from jet-drop to no-jet-drop regimes in the current experiments.

\begin{figure*}[!ht]
    \centering
    \includegraphics[width=1\linewidth]{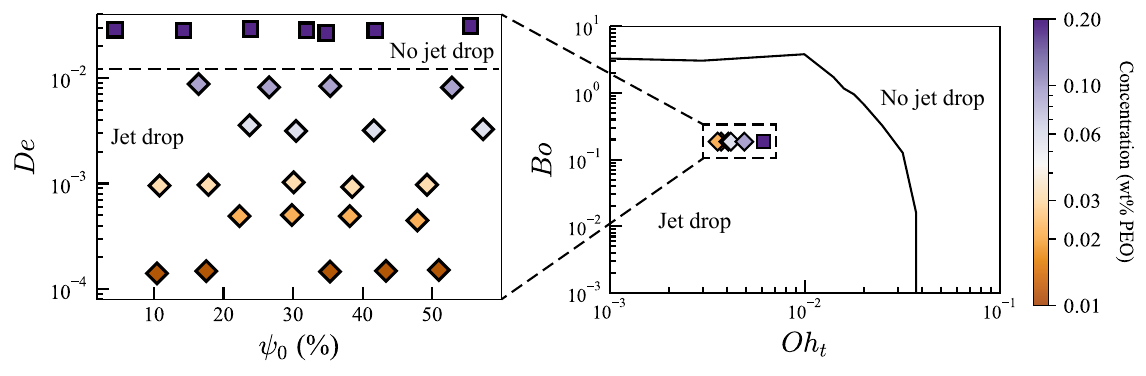}
    \caption{Regime maps depicting jet-drop and no-jet-drop transitions for jets from bursting bubbles with a viscoelastic compound layer (left) and bare bubbles (right). The left figure demonstrates the jet drop behavior in a $De-\psi_0$ space. Here, the diamond markers indicate the production of jet drops, the square markers indicate the absence of jet drops, and the dashed line indicates the experimentally observed regime boundary. The right figure demonstrates the projection of the experiments onto the parametric space of $Bo$ and $Oh_t$. The jet-drop and no-jet-drop regimes are reproduced from a previous study of bare bubble bursting$\cite{walls2015jet}$.}
    \label{fig:phase}
\end{figure*}

\section{Conclusion}
In conclusion, we experimentally investigated the dynamics of bubble bursting with a polymeric compound layer in a Newtonian fluid. By systematically varying the compound layer volume fraction and polymer concentration, we explored their impact on cavity collapse and jetting behavior. During the bursting of such a compound bubble, bubble cavity collapses with capillary waves focusing at the bottom, ejecting a jet that entrains the viscoelastic compound layer fluid into the atmosphere.  

We first find that the presence of the compound layer has negligible influence on the cavity collapse timescale, regardless of the polymer concentration or the compound layer volume fraction. Next, we observe a more slender jet with faster velocity and smaller radius when the compound layer fraction increases. We attribute the more energetic jet to the damping of capillary waves and the decrease of cavity cone angle before jet birth. Moreover, despite the nearly constant $Oh_t$ number across the polymer concentrations in our experiments, we observe a decrease in jet velocity and an increase in jet radius with increasing polymer concentration. This is due to the increasing viscoelasticity that introduces strong extensional stresses slowing down the stretching jet. When $De$ is smaller than 10$^{-2}$, the ejected jet breaks down into drops connected by elastocapillary filaments, which gradually thin due to tensile stretching. This process closely resembles the characteristic ``beads-on-a-string" structure for viscoelastic liquid thread thinning. Additionally, this viscoelastic filament exerts a drag force on the drops formed, decelerating their velocity more strongly at higher polymer concentrations with a longer polymer relaxation time. As the polymer concentration rises, $De$ increases and the number of drops decreases. The production of jet drops eventually ceases entirely at a polymer concentration of 0.2 wt\%. Additionally, the number of drops formed increases with the compound layer fraction for a given polymer concentration as the jet becomes more slender.  We also provide a regime map across a range of $De$ and $\psi_0$, illustrating the conditions under which jet drops are produced or suppressed in compound bubble bursting.

We believe that our findings advance the understanding of fluid mechanics and interfacial transport governing the bursting of bubbles coated with rheologically complex contaminants. Furthermore, this study offers valuable insights into the ocean-atmosphere mass transport of biochemical substances mediated by bubble bursting, which plays a critical role in marine biology and environmental science.

We believe that our findings advance the understanding of fluid mechanics and interfacial transport governing the bursting of bubbles coated with rheologically complex contaminants. Furthermore, this study offers valuable insights into the ocean-atmosphere mass transport of biochemical substances mediated by bubble bursting, which plays a critical role in marine biology and environmental science.
\section{Author contributions}

S.B.: Data curation; Investigation; Formal analysis; Methodology; Visualization; Validation; Writing – original draft; and Writing – review \& editing.
Z.Y.: Data curation; Investigation; Validation; and Writing –
review \& editing. J.F.: Conceptualization; Funding acquisition; Formal
analysis; Investigation; Project administration; Resources; Supervision; and Writing – review \& editing.

\section*{Conflicts of interest}

There are no conflicts to declare.

\begin{acknowledgement}

We thank the Materials Research Laboratory (MRL) at the University of Illinois at Urbana-Champaign for granting access to their stress-controlled rheometer for partial characterization of the polymer solutions. We also acknowledge Prof. Randy H. Ewoldt and Prof. Jonathan Freund at the University of Illinois at Urbana-Champaign for their valuable discussions and insights. S.B., Z. Y., and J.F. acknowledge partial support by the National Science Foundation (NSF) under grants No. CBET 2426809, 2323045 and the research support award RB24105  from the Campus Research Board at the University of Illinois at Urbana-Champaign.

\end{acknowledgement}




\bibliography{References}
\normalsize
\end{document}